\documentclass[prl,aps,twocolumn,showpacs,preprintnumbers,amsmath,superscriptaddress,floatfix,10pt]{revtex4}%{revtex4-1}

\usepackage{amsmath}
\usepackage[latin1]{inputenc}
\usepackage{bm}% bold math
\usepackage{graphicx}% Include figure files
\usepackage{SIunits}
\usepackage{subfigure}

\newcommand{\be}{\begin{equation}}
\newcommand{\ee}{\end{equation}}
\newcommand{\ba}{\begin{eqnarray}}
\newcommand{\ea}{\end{eqnarray}}
\newcommand{\ban}{\begin{eqnarray*}}
\newcommand{\ean}{\end{eqnarray*}}

\begin{document}

\title{\large{Experimentally faking the violation of Bell's inequalities}}

\author{Ilja Gerhardt}
\affiliation{The first two authors contributed equally to this study.}
\affiliation{Centre for Quantum Technologies, National University of Singapore, 3 Science Drive 2, Singapore 117543}
\affiliation{Chemistry Department, Low Temperature Group, University of British Columbia, Vancouver, B.C.\ V6T~1Z1, Canada}

\author{Qin Liu}
\affiliation{The first two authors contributed equally to this study.}
\affiliation{Department of Electronics and Telecommunications, Norwegian University of Science and Technology, NO-7491 Trondheim, Norway}

\author{Ant{\'i}a Lamas-Linares}
\affiliation{Centre for Quantum Technologies, National University of Singapore, 3 Science Drive 2, Singapore 117543}

\author{Johannes Skaar}
\affiliation{Department of Electronics and Telecommunications, Norwegian University of Science and Technology, NO-7491 Trondheim, Norway}
\affiliation{University Graduate Center, NO-2027 Kjeller, Norway}

\author{Valerio Scarani}
\affiliation{Centre for Quantum Technologies, National University of Singapore, 3 Science Drive 2, Singapore 117543}
\affiliation{Department of Physics, National University of Singapore, 3 Science Drive 2, Singapore 117543}

\author{Vadim Makarov}
\email{makarov@vad1.com}
\affiliation{Department of Electronics and Telecommunications, Norwegian University of Science and Technology, NO-7491 Trondheim, Norway}
\affiliation{University Graduate Center, NO-2027 Kjeller, Norway}

\author{Christian Kurtsiefer}
\email{christian.kurtsiefer@gmail.com}
\affiliation{Centre for Quantum Technologies, National University of Singapore, 3 Science Drive 2, Singapore 117543}
\affiliation{Department of Physics, National University of Singapore, 3 Science Drive 2, Singapore 117543}

\date{October 23, 2011}

\begin{abstract}
Entanglement witnesses such as Bell inequalities are frequently used to prove the non-classicality of a light source and its suitability for further tasks. By demonstrating Bell inequality violations using classical light in common experimental arrangements, we highlight why strict locality and efficiency conditions are not optional, particularly in security-related scenarios.
\end{abstract}


\maketitle

%% Intro, trust in measurement setups
\textbf{Introduction.} Experimental demonstrations of theoretical ideas require a reliable translation between the mathematical objects used in the theory and a specific setup of devices. In some instances trust may not be warranted due to the unnoticed failure of some measurement devices. In the context of quantum cryptography, a warning was raised recently: the possibility of so-called \emph{faked states}~\cite{makarov:05}. In this scenario, honest scientists are recording real measurement outcomes, performed by devices that seemingly work as they should. What is exploited is the fact that a physical device, even in its presumably normal state, may be sensitive to other degrees of freedom than the ones that are thought to be relevant. For instance, it is customary to encode qubits in (e.g., polarization) states of the light field; but the field is much more than a qubit and the ``rest" may trigger a detector as well.

%% Bell's inquality as an entanglement witness & loopholes
A parallel theoretical development, triggered by quantum cryptography in conjunction with foundational studies, proposes an unexpectedly powerful solution to the problem of trust using Bell's inequalities~\cite{bell:66}. These purely statistical criteria can be checked without any knowledge of the degree of freedom that is studied and of the measurements that are performed --- in fact, they are even independent of quantum physics. Therefore, whatever physical device leads to a \emph{violation of Bell's inequalities} can immediately be labeled as non-classical. This intuition can be refined to provide \emph{device-independent} criteria for secure cryptography~\cite{acin:06, acin:07}, trusted randomness~\cite{pironio:10}, entangled states~\cite{bardyn:09, bancal:11}, and measurements~\cite{rabelo:11}. These works, together with the parallel approach called \emph{self-testing}~\cite{mayers:04, mckague:11}, show that some quantum devices can be certified solely from the observed statistics. There is a warning, though: trust should be given if and only if the violation of Bell's inequalities is \emph{loophole-free}. In the context of Bell's inequalities, a loophole means that the observed statistics may in fact be due to classical communication (``locality loophole") or to shared randomness (``detection loophole"). If a loophole is not closed, the violation of Bell's inequalities can be produced by trivial classical means, for instance two suitably programmed and possibly causally connected computers.

%% general comments on loopholes and faked states. (warning)
It is widely accepted that the loopholes must be closed if the devices are not fully characterized. Experimentalists, however, assume to have an \emph{a-priori knowledge} that their data are not generated by a pair of conspiring computers, and may be tempted to adopt a more relaxed stance. A warning against this complacency was issued when an unplanned post-selection of data led to a violation of Bell's inequalities larger than theoretically possible~\cite{tasca:09}. Bell experiments, in which one photon is sent through an optical amplifier before detection~\cite{pomarico:11} are another instance where non-closed loopholes can lead to misinterpretation of results. Here we discuss another scenario: using faked-state techniques~\cite{makarov:05, makarov:09, lydersen:11, gerhardt:11}, we show a fake violation of Bell's inequalities \emph{in the same conditions in which ordinary tests with genuine entangled states are performed}, by exploiting the physics of single photon detectors. In particular, two of our experimental demonstrations feature no post-selection in a passive-choice scheme, and a third demonstration features active choices, with a post-selection that keeps half of the data (more than any Bell experiment reported to date). The take-away message is that, when dealing with a malicious adversary, it is necessary to close all the loopholes, even if the measurement devices are known to the user and seem to behave normally.

\begin{figure}
  \includegraphics[width = 8.6cm]{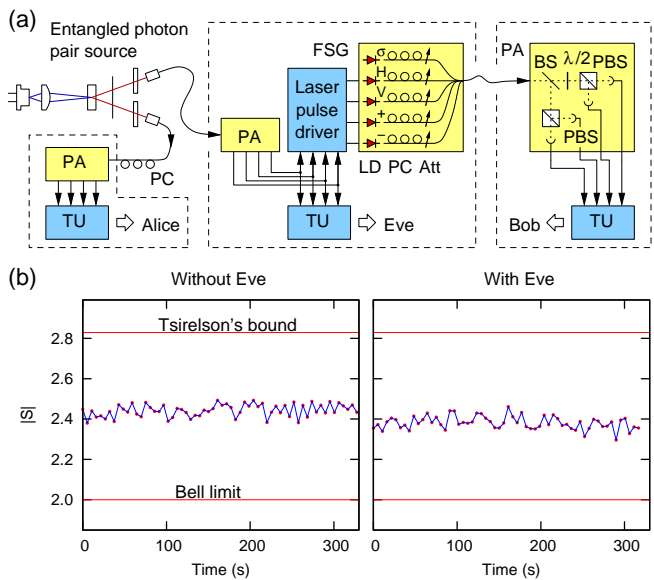}

  \caption{Testing a Bell inequality in an intercept-resend scenario. (a) Experimental setup. Pairs of polarization-entangled photons are generated via spontaneous parametric down-conversion (SPDC), and sent for polarization analysis (PA) in two conjugate linearly polarized bases on both sides using polarizing beamsplitters (PBS), half-wave plates ($\lambda/2$) and non-polarizing beamsplitters (BS); details are described in Ref.~\cite{marcikic:06}. An intercept-resend system (Eve) is inserted in the fiber line that carries one of the photons. Eve consists of a PA and faked-state generator (FSG), and generates measurement results that are copied into the receiver Bob. Key elements of the FSG are laser diodes (LD), polarization controllers (PC) and attenuators (Att); see details in Ref.~\cite{gerhardt:11}. Detection events are recorded with time-stamp units (TU) for later correlations. (b) From the correlations between the measurement results, Alice and Bob can test a CHSH Bell inequality. With the photodetectors used for this experiment, they obtain the same result (i.e.,\ a violation of a Bell inequality) with and without the presence of Eve, and would conclude that they witnessed entanglement in a pair of photons.}
  \label{fig:PDCsetup}
\end{figure}
 
%% the experimental Bell test, general remarks
\textbf{Experimental Bell test.} With a typical setup to generate entangled states we carry out polarization correlation measurements (Fig.~\ref{fig:PDCsetup}). A source of polarization-entangled photon pairs based on parametric down-conversion~\cite{kwiat:95} feeds each member of the pair to two legitimate parties, Alice and Bob, who measure the polarization in one of two possible bases using a beamsplitter as a random basis choice, and polarizers followed by avalanche single photon detectors (APDs).

%% a bit of math
The test measures photon polarization correlations between two parties in two bases,
 \begin{eqnarray}
 S &=& E_{AB}+E_{A'B}+E_{AB'}-E_{A'B'}\,,
 \end{eqnarray}
where $A, A'$ and $B, B'$ correspond to the measurements at Alice's and Bob's side respectively, and $E_{AB}$ is the correlation function
\begin{eqnarray}
E &=& \frac{P(1,1)-P(1,0)-P(0,1)+P(0,0)}{P(1,1)+P(1,0)+P(0,1)+P(0,0)}\,,
\end{eqnarray}
where $P$ is the probability of a joint outcome.
%% Intro to Bell test / max values.
The inequality~\cite{clauser:69} (CHSH-inequality) states that any classical state will have $|S| \leq 2$. Quantum mechanical states can produce up to $|S| \leq 2\sqrt 2$, and the mathematical maximum value is $|S|=4$. Note that this expression already post-selects only cases where pairs of detection events were seen on both sides --- it excludes the cases where one of the photons was lost due to inefficiencies in the detectors, source, or transmission path.

%% Bell's inequality with an eavesdropper, with SPDC source, introducing Eve
\textbf{Bell's inequality with an eavesdropper.} In between the source of entangled photons and Bob, we set up another observer, Eve. She has a measuring apparatus identical to Bob's, and a ``faked-state generator'' (FSG), which takes advantage of the detailed physical mechanism behind APDs to manipulate their output. When in normal operation, the arrival of a photon on an APD creates an electron-hole pair, which gets separated by an applied voltage and in the process creates an avalanche resulting in a macroscopic current; when the current exceeds a certain threshold, a ``click'' is registered and interpreted as the arrival of a photon. By sending tailored strong pulses of light into an APD, it is possible to manipulate the output such that ``clicks'' appear at Eve's will~\cite{makarov:09, gerhardt:11}: Beyond some optical power level, the detector does not have time to recharge and is constantly producing small avalanches that do not fulfill the threshold condition, and can thus be blinded or saturated. At even higher powers, the number of electron-hole pairs produced is such that the current can exceed the threshold condition set by the electronics without the need for an avalanche. By tailoring the polarization and power of the optical pulses sent into Bob's apparatus, it is possible to force him to obtain any desired measurement result.
 
\begin{figure}
\includegraphics[width = 7.6cm]{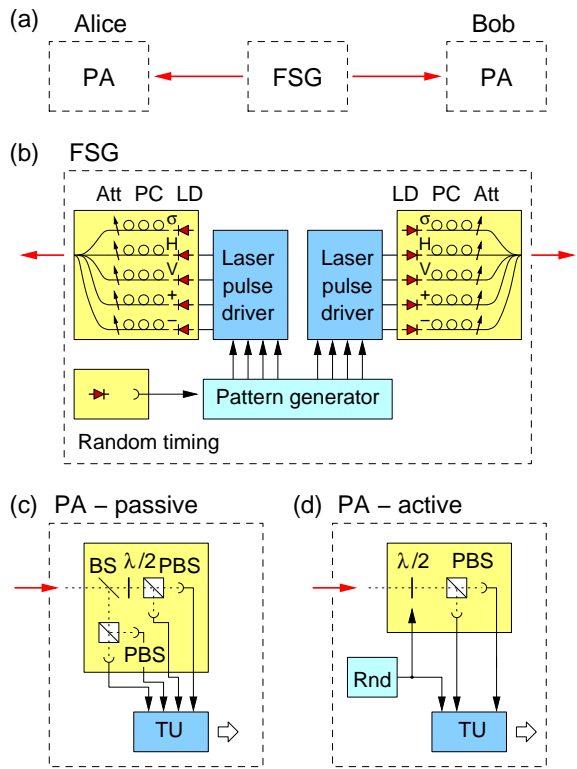}
  \caption{Bell test with the source replaced by a twin FSG. (a) Optical outputs of the FSG are connected to two polarization analyzers at Alice and Bob. (b) Implementation of the twin FSG: A pre-programmed list of correlation events is sent to two single FSGs at random times defined by a photocounting detector illuminated with a constant intensity. (c) Polarization analyzer with a passive choice of measurement basis, and (d) with an active choice, where a half wave plate ($\lambda/2$) is rotated by a motor according to the value of an independent random bit generator (Rnd).}
  \label{fig:setup}
\end{figure}
 
%% The FSG attack
An incoherent mixture of laser beams of two different polarizations allows the FSG (Fig.~\ref{fig:PDCsetup}(a)) to address any of the four detectors. One of the beams is circularly polarized ($\sigma$), thus gets distributed evenly among all detectors at a blinding intensity level. The other beam is linearly polarized in the same direction as the desired measurement result. The beamsplitter and polarization optics in the measurement setup will direct a larger fraction of the latter beam to the targeted APD than to any of the other APDs, driving only the targeted APD into the faking regime and forcing a false photon detection (see~\cite{gerhardt:11} for experimental details). The two parties testing the inequality would be registering ``clicks'' at the respective detectors and assembling joint probability distributions of the possible simultaneous results. While Alice is measuring genuine photons, Bob is looking at a classical pulse of light. A test of the CHSH inequality is shown in Fig.~\ref{fig:PDCsetup}(b). As this is an ``intercept-resend'' configuration, the Bell value registered will be a reflection of the entanglement quality of the original source. In this case the visibility in the $\pm$ basis is around $92\%$, and results in a value of $S=2.381\pm0.036$. This is well above the classical limit, highlights the possible breakdown of the Ekert quantum key distribution protocol~\cite{ekert:91}, and the need for strictly fulfilling the assumptions associated with the Bell theorem in a cryptography scenario.
 
%% a pre-programmed Bell violation, with no SPDC source
\textbf{Pre-programmed Bell violation.} Since we can fake any result, there is no need for the entanglement source, so we replace it with a pair of FSGs (Fig.~\ref{fig:setup}(a)). The polarizations expected by Alice are denoted $(\text{H},\text{V},+,-)$; those expected by Bob are rotated by $22.5\degree$ compared to Alice's and are denoted $(\tilde{\text{H}},\tilde{\text{V}},\tilde{+},\tilde{-})$. The frequencies $p_{P_AP_B}$ with which the FSG sends out polarization $P_A$ to Alice and $P_B$ to Bob are programmed to be \begin{eqnarray} \left(
\begin{array}{cccc}
p_{\text{H}\tilde{\text{H}}}&p_{\text{H}\tilde{\text{V}}}&p_{\text{H}\tilde{+}}&p_{\text{H}\tilde{-}}\\
p_{\text{V}\tilde{\text{H}}}&p_{\text{V}\tilde{\text{V}}}&p_{\text{V}\tilde{+}}&p_{\text{V}\tilde{-}}\\
p_{+\tilde{\text{H}}}&p_{+\tilde{\text{V}}}&p_{+\tilde{+}}&p_{+\tilde{-}} \\
p_{-\tilde{\text{H}}}&p_{-\tilde{\text{V}}}&p_{-\tilde{+}}&p_{-\tilde{-}}
\end{array}
\right) &=&
\left(
\begin{array}{cccc}
x&y&x&y\\ y&x&y&x\\ x&y&y&x \\ y&x&x&y
\end{array}
\right)\,,
\end{eqnarray}
where $x=(1+q)/16$, $y=(1-q)/16$, and $q$ varies from $-1$ to $+1$. Such correlations lead to a CHSH violation of $S=4q$: in particular, Tsirelson's bound~\cite{cirelson:80} is recovered for $q=1/\sqrt{2}$, while the choice $q=\pm1$ simulates the unphysical PR box~\cite{popescu:94}. 

\begin{figure}
\includegraphics[width=86mm]{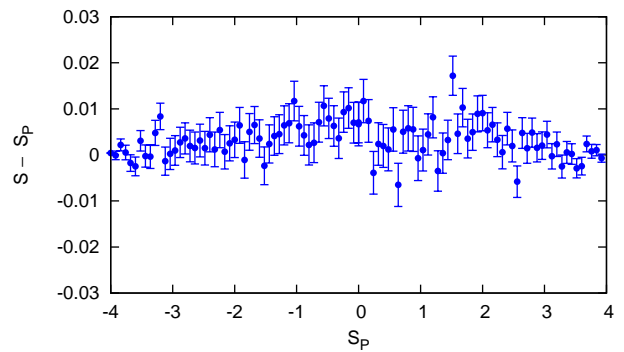}
  \caption{Difference between observed correlation function $S$ and expected $S_p$ in a CHSH Bell test, as a function of the programmed value $S_p=S(q)=4q$. The error bars were assigned through propagated Poissonian counting statistics.}
  \label{fig:sprogsobs}
\end{figure}

%% Experimental details, how FS are generated
For each value of $q$, a sequence of bright faked-state pairs is generated by a random number generator and the twin FSG (Fig.~\ref{fig:setup}(b)). Pairs are sent out upon photodetection events of an independent illuminated photodetector, to mimic the detection times of light emitted from the SPDC source. The experimental errors evaluated assuming Poisson counting statistics are on the order of $4\times10^{-3}$ when averaging over about 700 thousand events per $q$ value (Fig.~\ref{fig:sprogsobs}). Moreover, because the pulses of light can be made bright enough to overcome any channel losses, this configuration allows faking of correlations with efficiencies of $100\%$, meaning that there are no ``unpaired'' events: every time Alice registers a click, so does Bob.

%% question if active choice might help
The fixed nature of the analysis optics means that it is possible for the eavesdropper to use the same degree of freedom that is used to guide the genuine photons to the right detector (i.e.,\ polarization) in a malicious way to target her control pulses. Note that, although the random choice of measurement basis for an arriving photon is a genuine one~\cite{jennewein:00b}, its unchanging status results in an attack vector. So, is the problem solved just by having an active choice of basis? 

\begin{figure}
\includegraphics[width = 8.6cm]{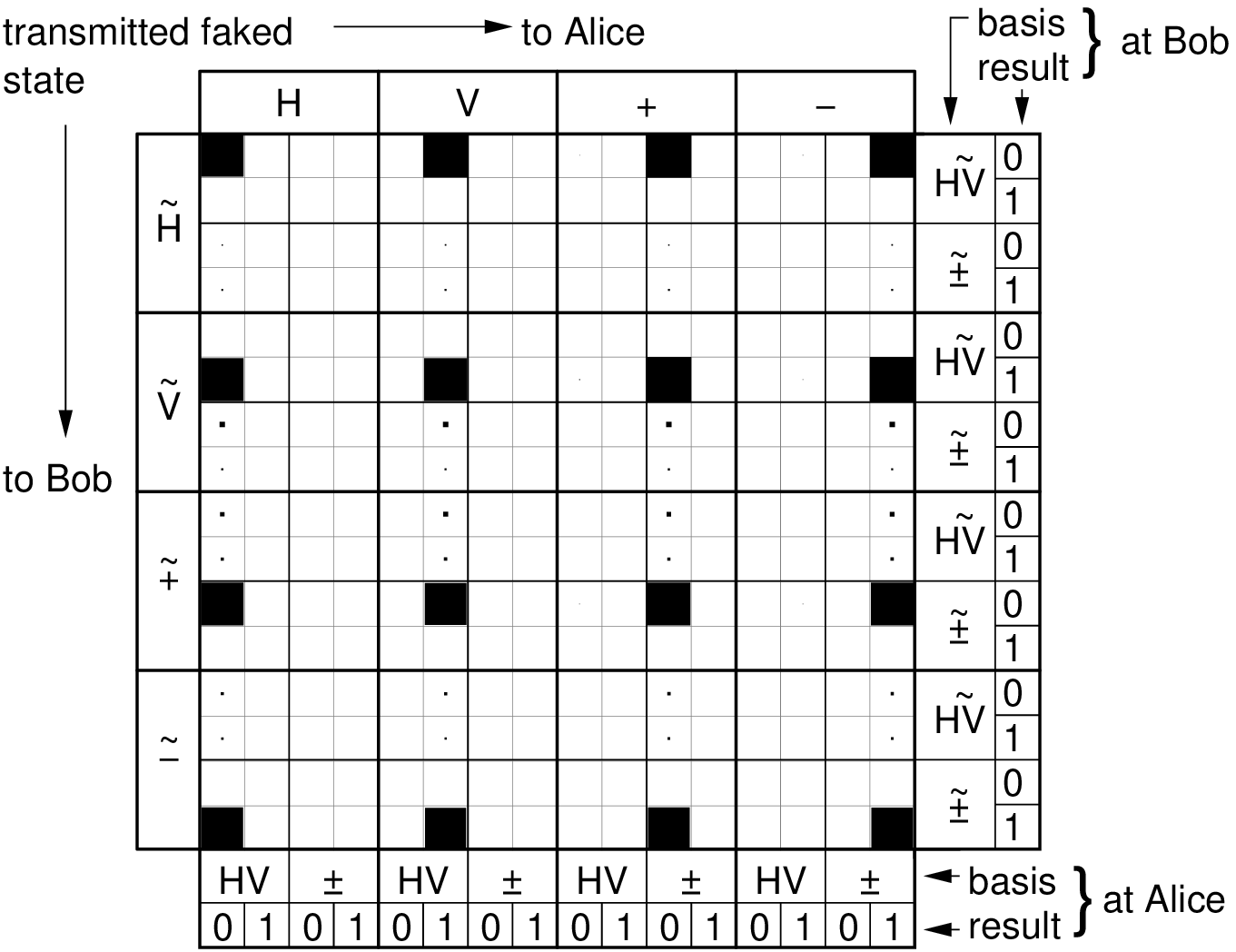}
  \caption{Experimentally obtained coincidences between pairs of detection events (0 or 1) at receivers Alice and Bob for their different measurement bases ($\text{HV}$ or $\pm$) as a function of different faked-state pairs of bright pulses sent to both sides. Each faked-state pair is made up by a combination of two pulses for horizontal ($\text{H}$) , vertical ($\text{V}$) and $\pm45\degree$ linear polarization ($+$ and $-$). Since bases for Alice and Bob are rotated in a Bell test by $22.5\degree$, we denote the faked-state polarizations and measurement bases on Bob's side with a tilde. Each cell's area is filled proportionally to the number of coincidence events; empty cell corresponds to 0 coincidence events, while a completely filled one corresponds to 40229 coincidences. The level of unwanted clicks is always below 1.4\% ($\leq 566$ coincidences).}
  \label{fig:correlation_matrix}
\end{figure}
 
%% Active basis choice introduction.
\textbf{Active basis choice.} We now come to the third experiment, where the measurement basis is chosen based on a random setting causally connected to neither the source nor the measurement apparatus. For this, we replace the BS by a half-wave plate that is rotated by a motor between two positions, determined by the parity of photodetection events from an illuminated, independent detector (see Fig.~\ref{fig:setup}(d)). The faking scheme will still work, once intensity levels are adjusted, but the immediate consequence is a decrease in efficiency from $100\%$ to $50\%$, since a basis choice not matched to the faked state results in an absence of the detection event. Efficiency here means number of paired detection events over total detected events, on each side. Although this reduction is dramatic, we are not aware of any experiment with entangled photons that has demonstrated an efficiency this high ($50\%$) so far. A correlation matrix between the prepared faked-state pairs, and the combination of detection events and measurement bases on both sides is shown in Fig.~\ref{fig:correlation_matrix}. Tiny imperfections in reproducing programmed coincidences (mean of unwanted clicks of 0.07\%) result from using a simplified version of the twin FSG, with a single bright-light source per polarization \cite{gerhardt:11}. Again, this matrix demonstrates full controllability of pair correlations even in an active basis choice scenario. Using this with a randomized distribution of faked coincidences corresponding to the Bell correlations, we obtain a value of $S=2.7971\pm0.0005$ in a run with 4561 basis settings (motor positions) and about $7.25\times 10^6$ coincidences. In principle, all values up to $|S|=4$ can be generated at will. Thus, an active choice by itself is still not sufficient to prevent a Bell inequality violation unless it is paired with an efficiency high enough to avoid the need for the fair sampling assumption.

%% Conclusion
\textbf{Conclusion.} The strength of Bell inequalities as entanglement witnesses is their formulation independently of the underlying physical model and measurement processes. Once we have an operational definition of what constitutes a measurement, such as the ``click'' mentioned repeatedly, there is a prescription to determine whether correlations between measurement results could be explained by a classical model. The results in this paper highlight how important the underlying assumptions in the theorem are to the interpretation of results. We have shown that ignoring the need for active choice and causal separation might result in an apparent violation with $100\%$ efficiency. Introducing an active choice scenario without consideration for the possible statistical bias is also not sufficient, and we achieve again an apparent violation beyond what has been achieved with genuine photonic entangled states. Our experiment shows that it is not legitimate to assume that a sampling of detection events is fair. Both loopholes need to be closed simultaneously for interpreting the violation of a Bell inequality as a refutation of local variables.

%%Countermeasures
While the detector control method used in this work could be defeated e.g.\ by monitoring optical power at the analyzer entrance \cite{lydersen:10,gerhardt:11}, such additional monitoring is neither a part of the Bell theorem nor of the security proofs for quantum cryptography. Furthermore, by trading off some detection efficiency, a control method requiring much less light (tens of $\pico\watt$) can be used \cite{makarov:09}, which would demand close to single-photon sensitivity of the power meter. The feasibility of implementing a countermeasure once the method of attack is explicit, does not detract from the conclusions, but rather highlights the need to carefully scrutinize any assumptions that are not part of the protocol.

%% Extension to other fields
We have focused here on Bell's inequalities, but it is evident that implementation of ``self-testing"~\cite{mayers:04, mckague:11} and even standard entanglement witnessing~\cite{skwara:07} needs some critical rethinking along the same lines.

We acknowledge useful discussions with Daniel Cavalcanti. This work was supported by the National Research Foundation and the Ministry of Education, Singapore, and the Research Council of Norway (grant no.\ 180439/V30).

%\bibliography{QIS-prl}

\end{document}